\documentclass[conference,a4paper]{APSIPA2020}
\usepackage{multirow}
\usepackage{graphicx}
\usepackage{amsmath}
\usepackage[psamsfonts]{amssymb}
\usepackage{amsxtra}
\usepackage{threeparttable}
\usepackage{epstopdf}
\usepackage{subfigure}
\usepackage[numbers,sort&compress]{natbib}

\begin{document}

\title{Attentive Fusion Enhanced Audio-Visual Encoding for Transformer Based Robust Speech Recognition}

\author{%
	\authorblockN{%
		Liangfa Wei, Jie Zhang, Junfeng Hou and Lirong Dai}
	\authorblockA{%
		National Engineering Laboratory for Speech and Language Information Processing \\
		University of Science and Technology of China (USTC), Hefei, Anhui, P. R. China \\
		E-mail: \{wlfxhgwt,hjf176\}@mail.ustc.edu.cn, \{jzhang6,lrdai\}@ustc.edu.cn}
}

\maketitle
\thispagestyle{empty}

\begin{abstract}
  Audio-visual information fusion enables a performance improvement in speech recognition performed in complex acoustic scenarios, e.g., noisy environments. It is required to explore an effective audio-visual fusion strategy for audio-visual alignment and modality reliability.
 Different from the previous end-to-end approaches where the audio-visual fusion is performed after encoding each modality, in this paper we propose to integrate an attentive fusion block into the encoding process. It is shown that the proposed audio-visual fusion method in the encoder module can enrich audio-visual representations, as the relevance between the two modalities is leveraged.
In line with the transformer-based architecture, we implement the embedded fusion block using a multi-head attention based audio-visual fusion with one-way or two-way interactions. The proposed method can sufficiently combine the two streams and weaken the over-reliance on the audio modality. Experiments on the LRS3-TED dataset demonstrate that the proposed method can increase the recognition rate by 0.55\%, 4.51\% and 4.61\% on average under the clean, seen and unseen noise conditions, respectively, compared to the state-of-the-art approach.
\end{abstract}

\section{Introduction}
Recently, with the rapid advance in deep learning, automatic speech recognition (ASR) has  become a reliable technique in high signal-to-noise ratio (SNR) environments. In case the SNR is too low, the ASR performance will degrade significantly, as the noise component dominates the microphone recordings. In order to alleviate the effects from noise and increase the  speech recognition performance, many efforts have been put on front-end techniques, e.g., speech quality and speech intelligibility enhancement~\cite{xu2014regression,zhang2018microphone,cees2013optimalSII}. However, the resulting speech recognition is still far from desired requirements.

In reality, apart from the audio observations, usually people also pay much attention to speaker's lip movements for better understanding the target speech, implying that human speech perception is bimodal in nature \cite{stork2013speechreading}. Inspired by this, the visual information can be leveraged to complement the traditional audio modality for speech recognition. The visual modality can become more important when the audio data is heavily contaminated by ambient noises.
It was shown that the video-based lip motion also contains discriminative speech content \cite{assael2016lipnet, stafylakis2017combining, wand2016lipreading}. The audio-visual speech recognition (AVSR) \cite{noda2015audio-visual, petridis2018end, afouras2019deep, petridis2018audio} that jointly uses the audio information and video information outperforms the traditional audio-only ASR system over a wide range of conditions, especially in noisy environments.
The combination of multi-modal information relies on suitable fusion strategies to model the audio-visual alignment and modality reliability. It is thus crucial to explore an effective fusion strategy for the AVSR, particularly on when and how to fuse the modalities.

With respect to the fusion stage, various audio-visual fusion strategies have been proposed~\cite{potamianos2003recent, katsaggelos2015audiovisual}. Generally, they can be classified into two categories: feature fusion~\cite{mroueh2015deep,teissier1999comparing} and decision fusion~\cite{teissier1999comparing, huang2013audio}. For the feature fusion, the audio features and the video features are fused in model and jointly utilized in decoding.
For the decision fusion, the audio-based and video-based speech recognition results are simply combined  to make a final decision, like ROVER \cite{fiscus1997post}. It was  shown that for AVSR systems, the feature fusion can achieve a better performance, as the relevance of the multi-modal features is taken into account \cite{petridis2018audio, zhang2019robust}. Moreover, recently the end-to-end AVSR  tends to perform audio-visual feature fusion on  higher-level representations of each modality after the encoding process. For example, a dual-attention mechanism \cite{chung2017lip,afouras2019deep} was included to combine the audio and visual representations in the decoder, which is designed to learn the correlation between different input modalities. In \cite{sterpu2018attention}, an improved fusion method was proposed to decouple each modality from the decoder and used an attention mechanism to fuse the audio and visual features on the top layer of the encoder.
However, separating the feature fusion from the input modality encoding as in previous approaches turns out to be sub-optimal (e.g., see section \ref{section5}).
Therefore, different from the previous end-to-end approaches, in this paper we integrate an attentive fusion block into the encoding process rather than after the encoding process, which can enrich the audio-visual representations by adopting the relevance between the two streams.
%We compare our feature fusion method on LRS3-TED dataset with that in decoder like \cite{afouras2019deep,chung2017lip} and that on the top layer of encoder like \cite{sterpu2018attention} and demonstrate that our method outperforms the previous method in both clean and noisy conditions for robust speech recognition.

Regarding how to fuse the audio-visual modalities, the most commonly-used method is concatenation \cite{afouras2019deep, petridis2018audio,chung2017lip,petridis2018end,zhang2019robust}. However, a simple feature concatenation  fails to learn the correlation between different modalities and the resulting model turns out to be strongly dependent on the audio modality.  Although the  attention-based mechanism in~\cite{sterpu2018attention} can automatically learn the alignment between the audio and visual modalities, which effectively enhances the speech representation,
%In~\cite{yu2020audio-visual}, based on voice separation, a gated fusion is used to implicitly separate overlapped voices for AVSR.
the enhanced audio stream was only sent to the decoder, which might not sufficiently explore the supplementary discriminative information from the video modality.
In order to better use the visual modality, in this paper, we follow the audio-visual concatenation and exploit a multi-head attention with one-way or two-way modality interactions to build an attentive fusion block. By integrating the fusion block into the encoding process, we propose an attentive fusion enhanced audio-visual encoding scheme for the transformer-based robust speech recognition.

The rest of this paper is organized as follows. In Section \ref{section2}, we introduce the structure of the proposed attentive fusion enhanced audio-visual encoding model and the existing feature fusion methods. In Section \ref{section3}, the embedded audio-visual fusion block is presented. Section \ref{section4} presents the experimental setup, including the database, audio and video features and the training strategy. Experimental results and discussions are given in Section \ref{section5}. Finally, Section \ref{section6} concludes this work.

\section{Deep fused Audio-visual Encoding}
\label{section2}
Feature fusion has been shown to be an effective and dominant method for the AVSR. However, the best stage to perform feature fusion is still unknown.
%The most common feature fusion methods recently are all carried out on the higher level representations which are performed after completing the encoding process of each modality such as in decoder or on the top layer of encoder.
Although the deep encoding of each modality can obtain high-level representation of the individual information, the subsequent fusion  to further model the relationship  between different modalities is difficult. In order to further utilize the coupling relationship between the audio and video data, we propose to integrate the feature fusion into the internal encoding module. For the convenience in following discussions, based on the fusion stage, we will refer the proposed method, fusion on the top layer of the encoder and fusion in the decoder to as Early-fusion, Middle-fusion and Late-fusion, respectively. Note that these three fusion strategies will be compared at the same model scale.
\subsection{Proposed Early-fusion encoding}
In this section, we will present the proposed deep fused audio-visual encoding architecture.
As a basic architecture, the transformer-based structure  \cite{vaswani2017attention} will be exploited for implementing the considered AVSR model. The transformer uses a self-attention mechanism to learn the long-term time dependence of speech features. It is characterized by the capability of directly calculating the dependency regardless of the distance between words. Compared to the recurrent neural network (RNN) that requires an inherent order, it has better parallelism and batch processing capability for long sequences.

 The detailed structures of the proposed Early-fusion encoder and the decoder are depicted in Fig. 1(a) and Fig. 1(c), respectively.
At first, the two modalities are encoded separately using three transformer encoding blocks. Each block consists of a multi-head attention (MHA) and a feed forward network (FFN). Then, we combine the two modal features by an embedded fusion block to obtain the concatenated representation in the fusion domain. We deeply encode the fused features to simultaneously model the audio and video information and the interrelationship between them. Finally, we use a decoder for a single stream to recognize the spoken speech. Each decoder block takes the encoder output and the previous block's output (or, for the first block, the prediction from the previous decoding step of the network) as the input. Since the encoder acts on the representative fused features, the model can sufficiently combine the two streams and learn the implicit relationship between the two modalities. Meanwhile, due to the fact that the video information is combined with the audio features by the fusion encoding, the over-reliance on the audio modality can be weakened.

\subsection{State-of-the-art fusion encoding}
\subsubsection{Middle-fusion}
To compare the proposed encoding scheme to the Middle-fusion and the Late-fusion, we use the same transformer based architecture to implement the feature fusion on the top layer of the encoder. The  structure of the encoder is shown in Fig. 1(b) and the decoder keeps the same as that in Early-fusion.
For the Middle-fusion method, the fusion block is performed after the audio and video features are fully  separately encoded, such that the higher level representations of each modality are then obtained.
%If the fusion block is a one-way interaction which will be introduced in Section 3, the model is then similar to~\cite{sterpu2018attention}. The only difference lies in that we use an additional audio-visual concatenation to supplement the discriminative video information.

\begin{figure}[!t]
\centering
\subfigure[Early-fusion encoder]{
\includegraphics[width=0.33\linewidth]{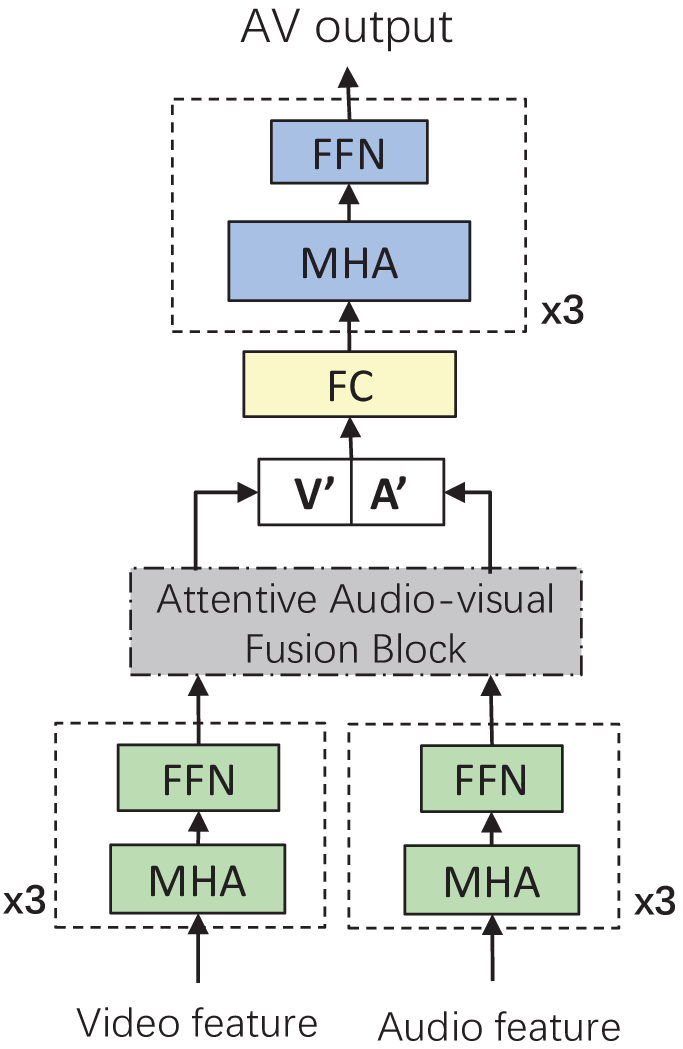}}%
\subfigure[Middle-fusion encoder]{
\includegraphics[width=0.33\linewidth]{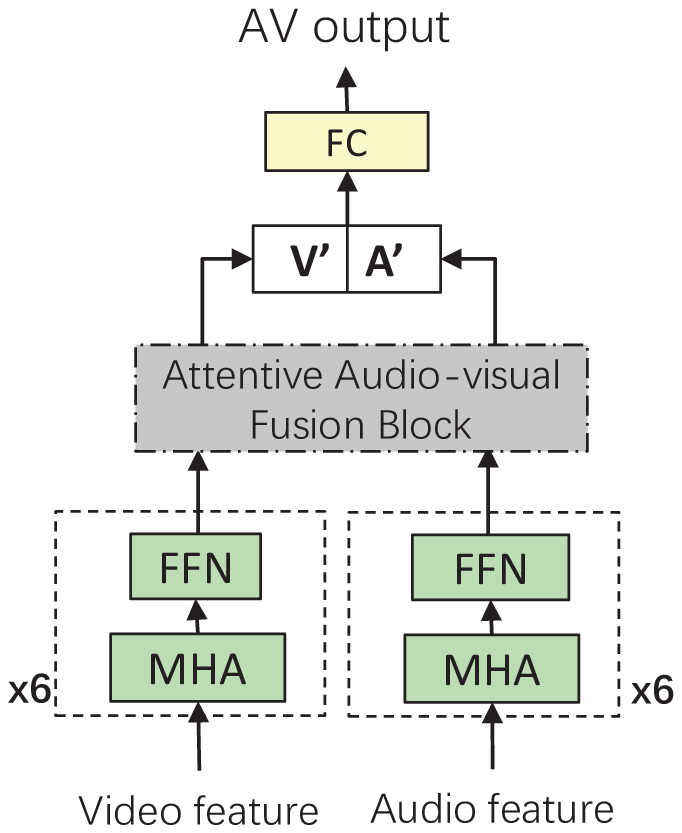}}%
\subfigure[decoder]{
\includegraphics[width=0.33\linewidth]{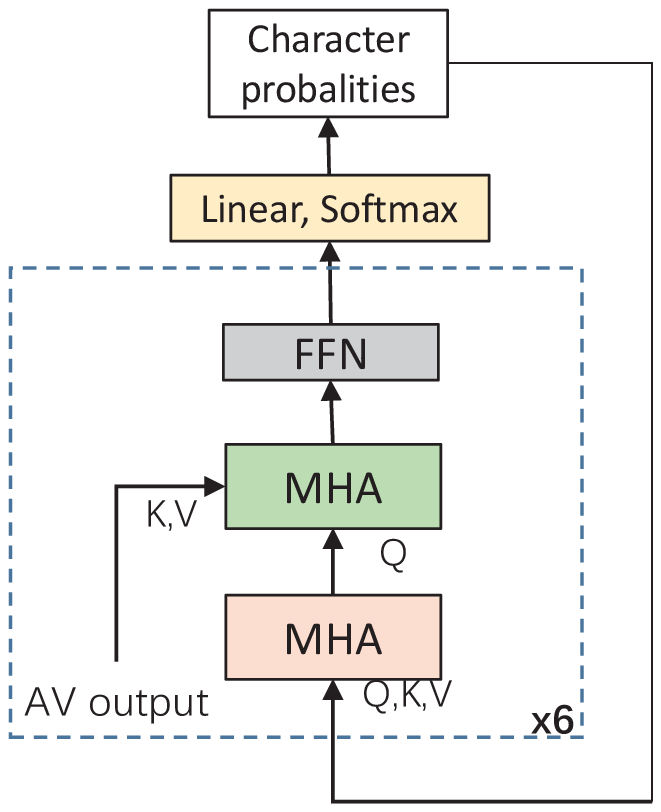}}%
%\end{center}
\caption{The AVSR architecture using early/middle feature fusion strategy. The involved MHA  takes 16 heads into account.  The size of the input layer, hidden units and the output layer of the FFN is 512, 2048 and 512, respectively.  The output dimension of the fully connect (FC) layer after the audio-visual concatenation is 512.
}
\end{figure}

\subsubsection{Late-fusion}
The Late-fusion strategy, as shown in Fig. 2, originates from~\cite{afouras2019deep}. The Late-fusion uses the dual attention mechanism to couple the audio and video features at the decoder side. In addition, We add an attentive fusion block after the encoder, such that it can be more fairly compared with the Early-fusion and Middle-fusion methods.

\begin{figure}[!t]
\centering
\subfigure[Late-fusion encoder]{
\includegraphics[width=0.4\linewidth]{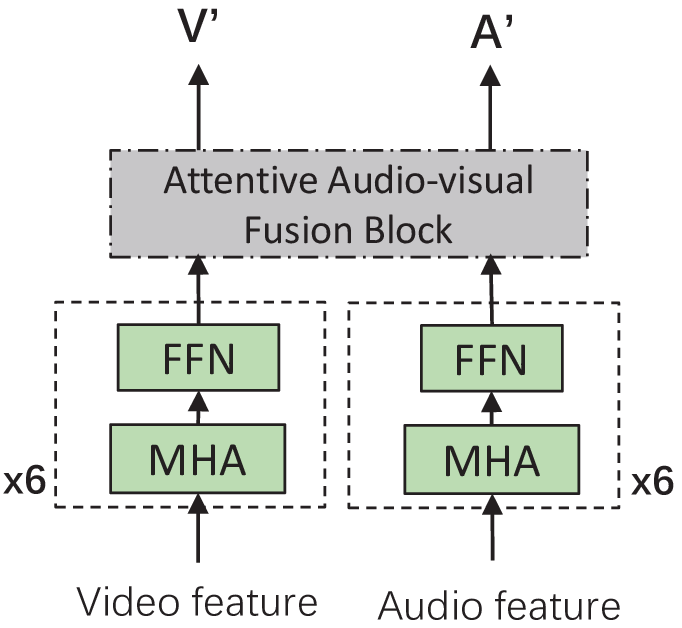}}%
\subfigure[Late-fusion decoder]{
\includegraphics[height=35mm,width=0.4\linewidth]{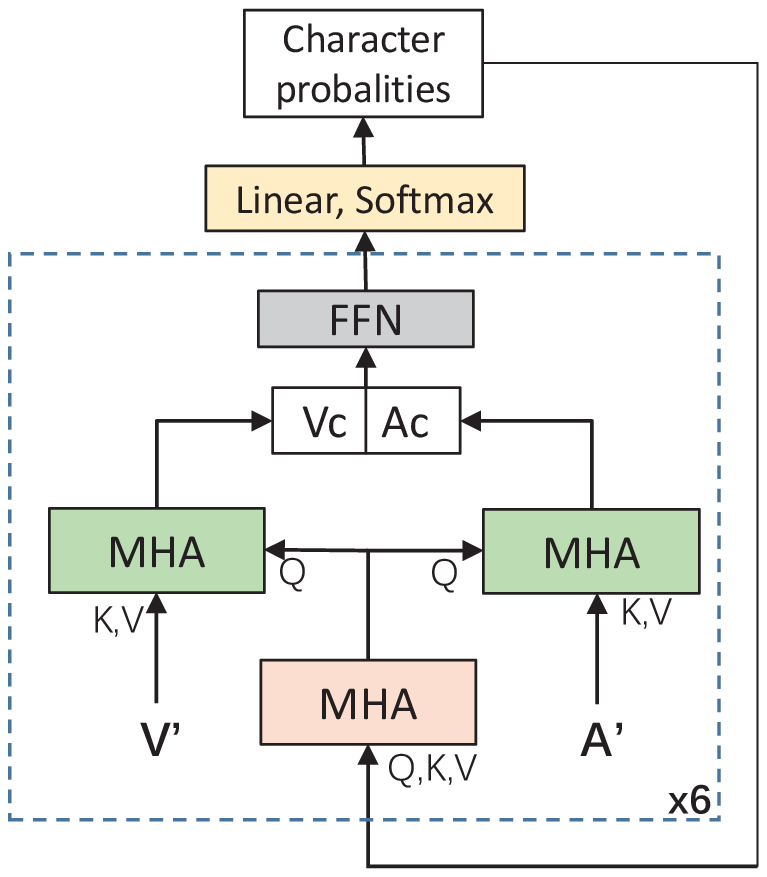}}%
%\end{center}
\caption{The AVSR architecture based on late feature fusion.}
\end{figure}

\section{Attentive Audio-visual Fusion Block}
\label{section3}
In this section, we consider to use an MHA based audio-visual (AV) fusion block for the proposed AVSR architecture.
In order to explore an effective fusion for the proposed encoding method, we will take the one-way and two-way interactions into account. The one-way MHA based fusion, which allows the audio modality to aggregate information from the video,, is referred to as AV-align. The two-way interaction, which allows the audio and video to aggregate information from each other, is referred to as AV-cross. The concatenation without fusion block is referred to as AV-concat.

\begin{figure}[!t]
\centering
\subfigure[AV-align (One way)]{
\includegraphics[height=45mm,width=0.22\textwidth]{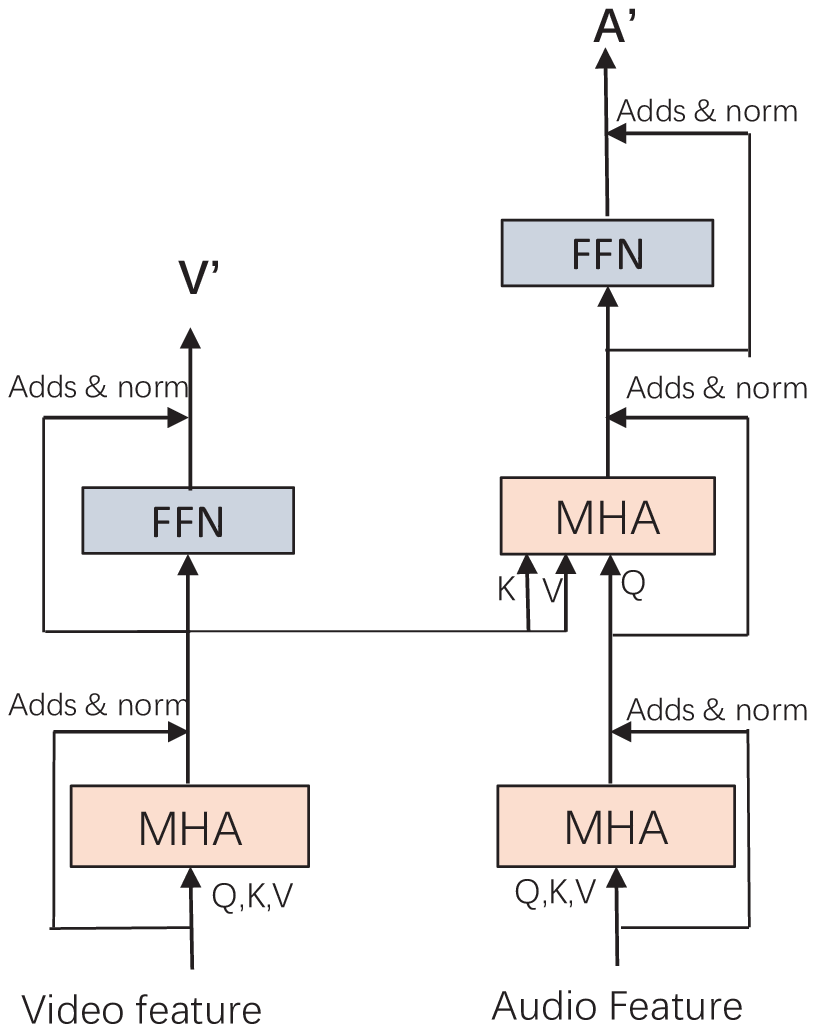}}%
\subfigure[AV-cross (Two way)]{
\includegraphics[height=45mm,width=0.22\textwidth]{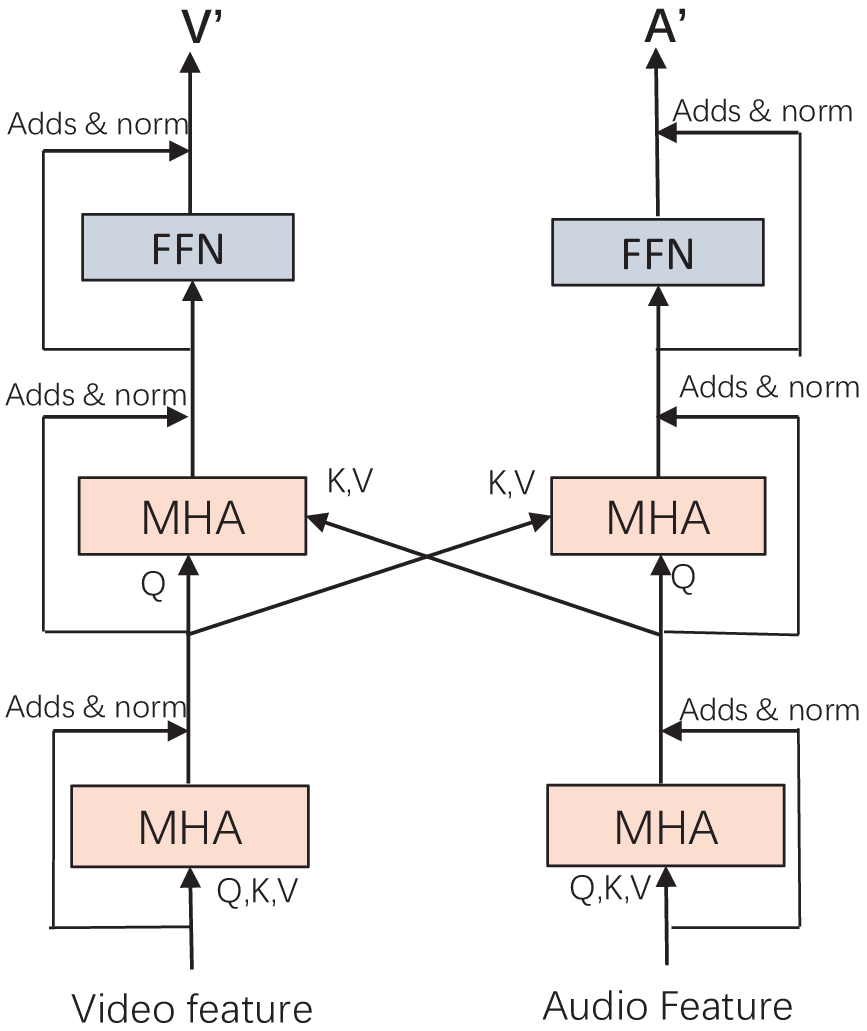}}%
%\end{center}
\caption{The attentive fusion block with one/two-way interactions}
\end{figure}

\subsection{AV-align}
In order to better correlate the acoustic features with the visual features at each time step, we use an AV-align block, which is shown in Fig. 3(a). The input vector of the MHA includes three parts: the input query (Q) vector extracted from the acoustic features after modeling, the input key (K) and value (V) vectors from the visual features.
 As such, the attention mechanism latently adapts streams from the audio modality to the visual modality, in which the features are aligned. We then add the attended video features to the input acoustic vector. The motivation behind this is to enable the acoustic modality to learn the synchronization information from the visual modality, and then supplement the original audio features to obtain an enhanced representation with more information and better noise robustness.

\subsection{AV-cross}
For AVSR, as usually the audio data contains far more information on the target speaker than the video observations, the model always tends to rely more on the audio features. The AV-align incorporates the audio-video synchronization information for the audio representation, which improves the utilization of the visual modality to a certain extent. To further resolve the over-dependence on the audio modality, we explore a two-way interaction for the audio-visual fusion, which is shown in Fig. 3(b). Clearly, the depicted AV-cross has a symmetric structure. Specifically, in contrast to the AV-align (where the audio features is taken as the Q vector, video features as the K and V vectors for the audio-side MHA), the AV-cross also takes the audio features as the K and V vectors and the video features as the Q vector for the video-side MHA. Therefore, we can obtain an enhanced visual representation, which implicitly includes the synchronization information from the audio modality. The information gap between the two modalities can thus be reduced, and  the model can then balance the utilization of the two modalities.

\section{Experimental Setting}
\label{section4}
To verify the proposed method, we will present the experimental setup, including the audio-visual database, dual-modal feature extraction and the training procedure in this section.

\begin{figure*}[t]
\centering
\subfigure[AV-concat]{
\includegraphics[width=0.3\textwidth]{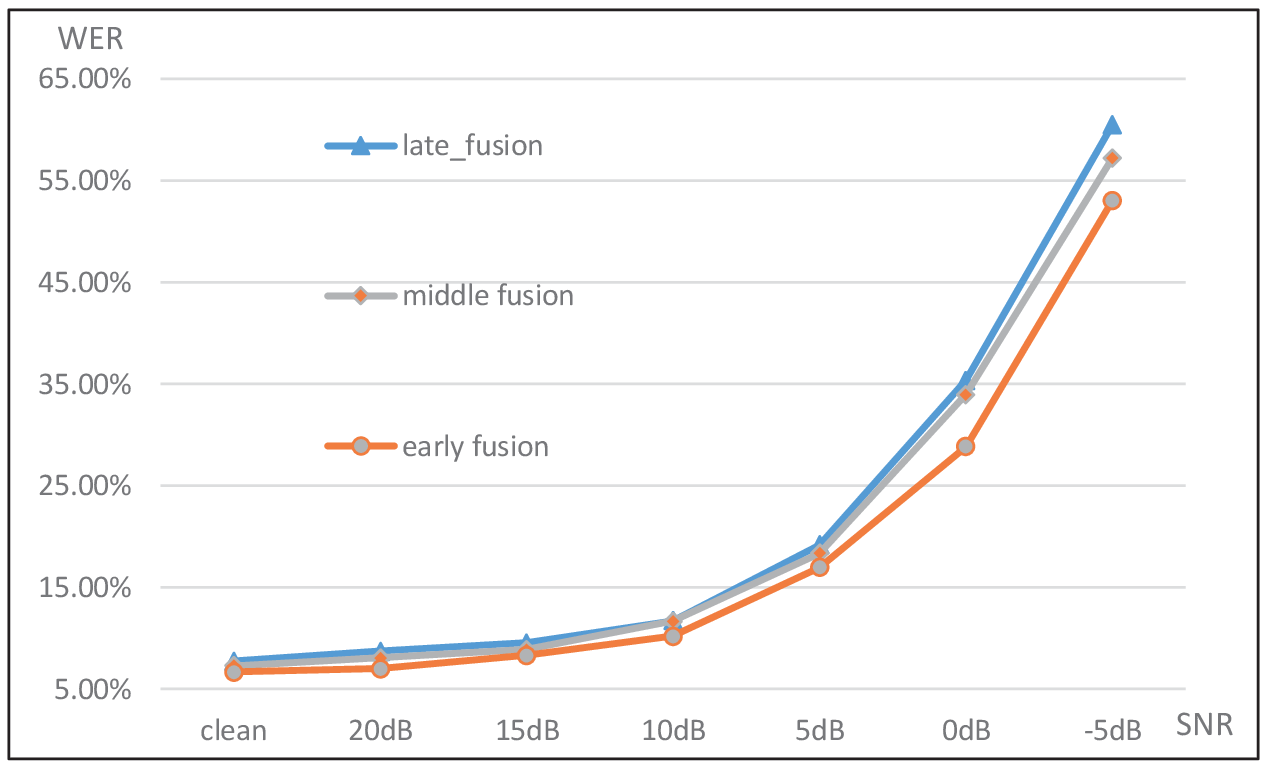}}%
\subfigure[AV-align]{
\includegraphics[width=0.3\textwidth]{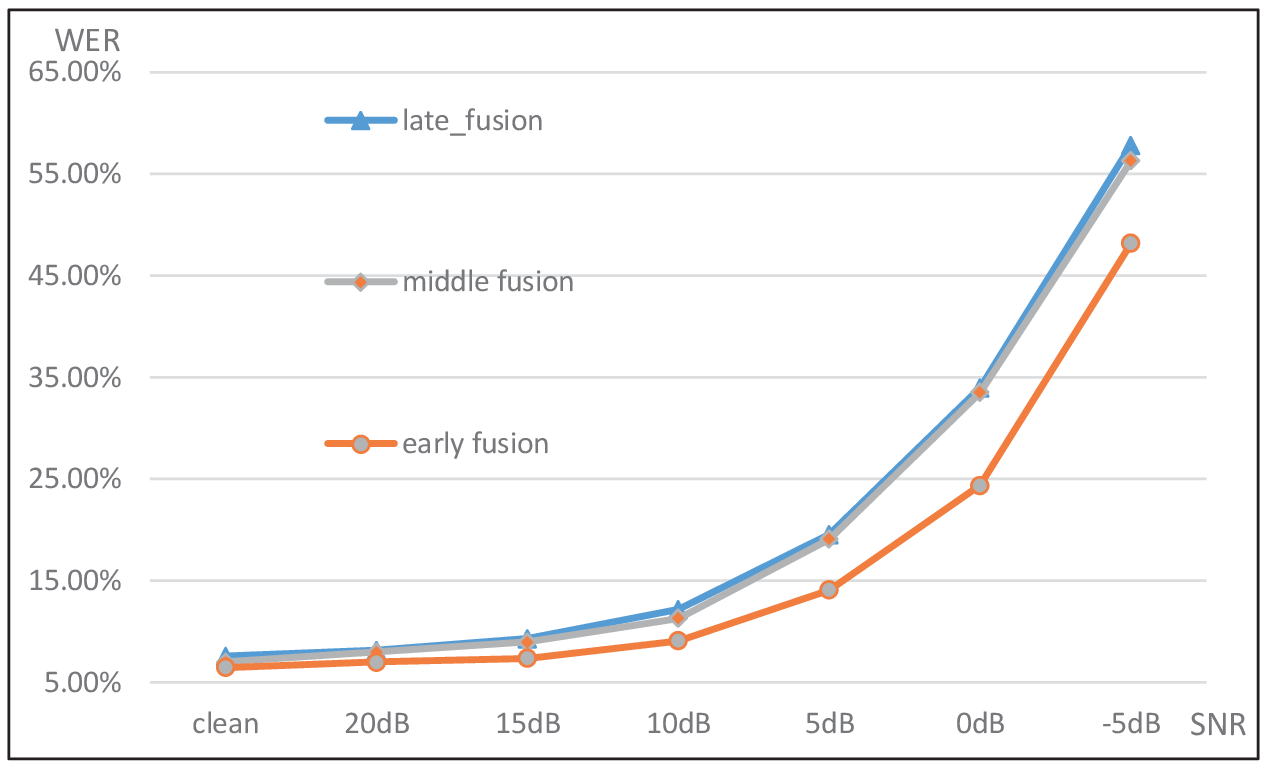}}%
\subfigure[AV-cross]{
\includegraphics[width=0.3\textwidth]{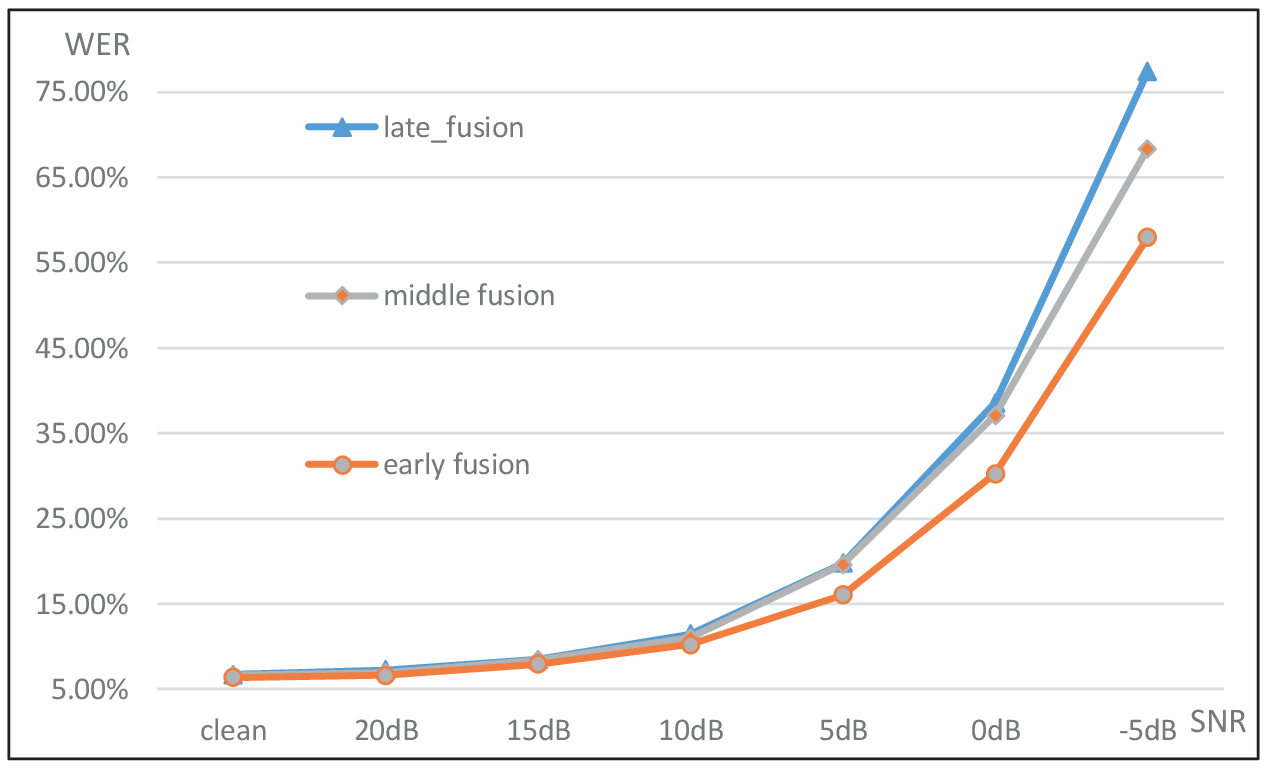}}%
%\end{center}
\caption{The WER in terms of the SNR using the babble noise for different fusion blocks: (a) AV-concat, (b) AV-align and (c) AV-cross.}
\label{figure:wer-curves}
\end{figure*}

\subsection{Database}
The proposed AVSR method is evaluated on the  LRS3-TED dataset \cite{afouras2018lrs3}, which is one of the largest public audio-video datasets in publicity. It contains 5594 videos downloaded from YouTube and over 400 hours video clips extracted from TED English speeches and TEDx speeches. The dataset is divided into three groups: a 444-hour pre-training set, a 30-hour training set, and a 1-hour testing set, which contain 132k, 32k, and 1452 samples, respectively. The pre-training set and the training set are extracted from the same set of YouTube video species, so they overlap in content, but the testing set is completely independent.
%The pre-training set is mostly long sentences, and the average length of the training set and the test set samples is equivalent.
For the model pre-training, in addition to LRS3-TED, we also use the pre-training set from LRS2-BBC \cite{afouras2019deep} to help the model obtain a good initial recognition ability.

For robust speech recognition, we regard the audio contained in LRS3-TED as the clean audio, as they are of high quality. To synthesize the noisy sensor observations, we  choose {white, babble, pink, factory1, factory2, and volvo} noises from the Noisex-92 dataset \cite{varga1993assessment} and add them to the LRS3-TED training set at different SNR levels. During the testing phase,  we use the babble noise as the seen noise and the buccaneer2 noise as the unseen noise, respectively.

\subsection{Audio and video features}
\textbf{Video Feature:} The video provided by LRS3-TED has a resolution of 224 $\times$ 224 at a frame rate of 25 fps. Since the mouth ROI is already centered, we directly resize the image of each frame into 112 {$\times$}112 pixels. Then, we use a 3DCNN and Resnet18 \cite{he2016deep} in combination  with a 512-dimensional FC layer as the visual front to extract the frozen feature vector. Similarly to \cite{afouras2018deep}, the parameters of the visual front are pre-trained as a word-level lip-reading task.

\textbf{Audio Feature:} We compute an 80-dimensional filter-bank-wise feature with a 25 ms window and 10 ms stride at a sampling rate of 16 kHz. Since the frame rate of the video is 25 fps, one video frame corresponds to 40 ms audio. In order to obtain the same temporal scale for both modalities, we concatenate the audio features in groups at an order of 4. That is, we extract a 320-dimensional audio feature vector, as the video frame rate is four times the audio shift.

\subsection{Multi-condition training}
For model training, we  first select the samples less than 1500 audio frames (limited by the GPU memory) from the clean pre-training set to pre-train the model. In order to make the trained model robust against noise, we consider multiple noise conditions characterized by different SNR levels. Specifically, we add the six types of noises mentioned in Section 4.A at an SNR level in \{20, 15, 10, 5, 0, -5\} dBs to the LRS3-TED training set. Then, we uniformly randomly select a clean speech or a noisy speech (with the six SNR levels and six noise types being taken into account) as each training sample.

During the pre-training phase, we introduce the curriculum learning (CL) \cite{bengio2009curriculum} in the first 4 epochs. During the formal training phase, we set 2-epoch CL at the beginning. The learning rate gradually decreases from $1^{-4}$ to $5^{-5}$ and from $1^{-4}$ to  $5^{-6}$ for pre-training and training phases, respectively. Teacher forcing, label smoothing and dropout with $p = 0.1$ are used for training. For testing, a beam search with a width of 6 is used.

\section{Experiment Results and Discussion}
\label{section5}
\subsection{Pre-experiment}
In order to verify the effectiveness of the multi-condition training strategy and the benefits of feature concatenation, we compare the proposed late fusion model without the attentive fusion block to \cite{afouras2019deep}, where the babble noise is added to the audio stream at a SNR of 0 dB with a probability of $p = 0.25$. Besides, in~\cite{sterpu2018attention}, a merely AV-alignment based AVSR approach was proposed, where the video stream is only used for aligning the audio data and the model only sends the audio stream to the decoder. To observe the superiority of the AV concatenation, we will compare the AV-align enhanced middle fusion model, which also uses the AV concatenation, to the method in \cite{sterpu2018attention}.
The word error rates (WERs) of the considered AVSR system under the clean and babble noise conditions are shown in Table~\ref{table:wers-noise}. It can be seen that using the proposed multi-condition training can reduce the WER by 0.24\% and 8.96\% under the clean and 0 dB babble noise conditions, respectively. Therefore, we will use the multi-condition training uniformly in the sequel.
Further, the AV-align enhanced middle fusion model can decrease the WER by 0.70\%, 1.95\%, and 11.25\% in the case of clean, 0 dB, and -5 dB SNRs, respectively, compared to the AV-align without concatenation. Hence, we can conclude that the feature concatenation is beneficial for improving the AVSR performance.
%So we replace the traditional AV-align method in \cite{sterpu2018attention} with our AV-align and concatenation.
Since the middle fusion is broadly used and the way of attention based AV-align fusion is recently proposed \cite{sterpu2018attention}, the AV-align enhanced middle-fusion will be taken as the baseline system.

\begin{table}[!t]
\begin{center}
\begin{threeparttable}
\caption{The WERs of the considered AVSR system under the clean and babble noise conditions.}
\begin{tabular}{|c|c|c|c|c|}
\hline
    Train strategy & Model &  clean & 0dB & -5dB \\
\hline\hline
    p=0.25 noisy train & Late-fusion \cite{afouras2019deep} &  8.00\% & 44.30\% & -- \\
\hline
    \multirow{3}*{multi-condition train} & Late-fusion &  7.76\% & 35.34\% & 60.50\% \\
\cline{2-5}
       & AV-align as \cite{sterpu2018attention} &  7.69\% & 35.44\% & 67.49\% \\
\cline{2-5}
       & AV-align+concat &  6.99\% & 33.49\% & 56.24\% \\
\hline
\end{tabular}
\label{table:wers-noise}
%\begin{tablenotes}
%\item[a] Uppercase
%\end{tablenotes}
\end{threeparttable}
\end{center}
\end{table}

%And \cite{afouras2019deep} also used a large non-public data set MV-LRS for pre-training, so the amount of audio and video data used is larger than that we use.

\subsection{Attentive Fusion Enhanced Audio-visual Encoding}

In this subsection, we  train the proposed Early-fusion model, Middle-fusion model and Late-fusion model,
%using different audio-visual fusion strategies,
where the training setup keeps the same. We consider two types of noises, i.e., the babble noise and bulcanneer2 noise. The babble noise is involved in the training phase, while the bucanneer2 noise is not included for training. Both noises are added to the clean data at different SNR levels.

\begin{table*}[!t]
\begin{center}
\begin{threeparttable}
\caption{A comparison of WERs using different fusion timing strategies and fusion blocks under multiple Babble noise conditions.}
\begin{tabular}{|c|c|c|c|c|c|c|c|c|c|}
\hline
    Fusion stage & Fusion Block &  clean & 20dB & 15dB & 10dB & 5dB & 0dB & -5dB & mean on noisy data \\
\hline\hline
      \multirow{3}*{Late-fusion} & AV-concat & 7.76\% & 8.75\% & 9.56\% & 11.72\% & 19.17\% & 35.34\% & 60.50\% & 24.17\% \\
\cline{2-10}
                                 & AV-align  & 7.54\% & 8.13\% & 9.29\% & 12.13\% & 19.52\% & 33.92\% & 57.79\% & 23.46\% \\
\cline{2-10}
                                 & AV-cross  & 6.73\% & 7.25\% & 8.48\% & 11.47\% & 19.76\% & 38.55\% & 77.38\% & 27.15\% \\
\hline\hline
    \multirow{3}*{Middle-fusion} & AV-concat & 7.32\% & 8.06\% & 8.88\% & 11.70\% & 18.37\% & 33.96\% & 57.22\% & 23.03\% \\
\cline{2-10}
                   & AV-align (baseline) & 6.99\% & 7.99\% & 8.94\% & 11.30\% & 19.08\% & 33.49\% & 56.24\% & 22.84\% \\
\cline{2-10}
                                 & AV-cross  & 6.64\% & 6.93\% & 8.42\% & 11.03\% & 19.60\% & 37.08\% & 68.35\% & 25.24\% \\
\hline\hline
     \multirow{3}*{Early-fusion} & AV-concat & 6.67\% & 7.02\% & 8.33\% & 10.19\% & 16.96\% & 28.84\% & 53.03\% & 20.73\% \\
\cline{2-10}
                                 & AV-align  & 6.44\% & 6.97\% & \textbf{7.34\%} & \textbf{9.09\%} & \textbf{14.07\%} & \textbf{24.37\%} & \textbf{48.16\%} & \textbf{18.33\%} \\
\cline{2-10}
                                 & AV-cross  & \textbf{6.39\%} & \textbf{6.64\%} & 7.96\% & 10.21\% & 16.04\% & 30.25\% & 57.94\% & 21.51\% \\
\hline
\end{tabular}
%\begin{tablenotes}
%\item[a] Uppercase
%\end{tablenotes}
\label{table:wers-babble}
\end{threeparttable}
\end{center}
\end{table*}

\begin{table*}[t]
\begin{center}
\begin{threeparttable}
\caption{The WERs of the considered AVSR models using the unseen \textbf{buccaneer2} noise at different SNR levels.}
\begin{tabular}{|c|c|c|c|c|c|c|c|c|c|}
\hline
    Fusion stage & Fusion Block & clean & 20dB & 15dB & 10dB & 5dB & 0dB & -5dB & mean on noisy data\\
\hline\hline
      \multirow{3}*{Late-fusion} & AV-concat & 7.76\% & 9.61\% & 11.53\% & 14.49\% & 20.94\% & 33.42\% & 50.99\% & 23.50\% \\
\cline{2-10}
                                 & AV-align  & 7.54\% & 9.61\% & 10.78\% & 14.40\% & 20.57\% & 32.41\% & 48.46\% & 22.71\% \\
\cline{2-10}
                                 & AV-cross  & 6.73\% & 8.76\% & 10.25\% & 13.63\% & 20.90\% & 35.93\% & 62.82\% & 25.38\% \\
\hline\hline
    \multirow{3}*{Middle-fusion} & AV-concat & 7.32\% & 9.20\% & 10.60\% & 14.08\% & 20.12\% & 31.29\% & 48.28\% & 22.26\% \\
\cline{2-10}
                   & AV-align (baseline) & 6.99\% & 8.88\% & 10.58\% & 13.65\% & 20.06\% & 30.85\% & 48.55\% & 22.10\% \\
\cline{2-10}
                                 & AV-cross  & 6.64\% & 8.47\% & 10.10\% & 13.81\% & 20.43\% & 34.07\% & 55.95\% & 23.81\% \\
\hline\hline
     \multirow{3}*{Early-fusion} & AV-concat & 6.67\% & 8.28\% & 10.12\% & 12.98\% & 17.81\% & 26.67\% & 41.89\% & 19.63\% \\
\cline{2-10}
                                 & AV-align  & 6.44\% & \textbf{7.86\%} & \textbf{9.05\%} & \textbf{10.39\%} & \textbf{15.04\%} & \textbf{23.79\%} & \textbf{38.80\%} & \textbf{17.49\%} \\
\cline{2-10}
                                 & AV-cross  &  \textbf{6.39\%} & 7.99\% & 9.61\%  & 12.23\% & 17.69\% & 28.42\% & 45.76\% & 20.28\% \\
\hline
\end{tabular}
%\begin{tablenotes}
%\item[a] Uppercase
%\end{tablenotes}
\label{table:wers-buccaneer2}
\end{threeparttable}
\end{center}
\end{table*}

In Fig.~\ref{figure:wer-curves},  we show the WER curves for the early-fusion, middle-fusion and late-fusion methods using three fusion blocks.
It is clear that regardless of the fusion block that is used by the model, the proposed Early-fusion method can always obtain the best performance under any noisy condition, while the Late-fusion method is always the worst. As pointed out in \cite{sterpu2018attention}, the Late-fusion strategy adds another attention mechanism that attends to the second modality, and requires the decoder to also learn  the correlation between the input modalities, which might overload the decoder. The Middle-fusion method avoids the overload  problem, but it cannot  accurately model the connotation information between the audio and video data, since the two streams are fused after a completely separate modeling and directly fed into the decoder. Compared to the Late-fusion and the Middle-fusion method, the proposed Early-fusion method can thus not only avoid the overload problem, but also combine the two streams more deeply. The deep combination of modalities enriches the audio-visual representations, as the relevance between the two streams is implicitly incorporated.
%The best performance of our Early-fusion encoding is 6.39\% WER using AV-cross fusion block in clean condition and 18.33\% WER on average using AV-align fusion block in noisy condition, which yields 8.58\% and 19.75\% relative WER reduction from the optimal baseline method.

In order to investigate an appropriate encoding strategy, we compare the AVSR performance using different fusion blocks in Table~\ref{table:wers-babble}. Note that this is a similar comparison to Fig.~\ref{figure:wer-curves}, but from a different perspective. Obviously, in case the fusion time is fixed, in general the AV-align performs better than the AV-concat under both the clean and noisy conditions.
The AV-cross reaches a lower WER than the AV-align under high SNR conditions, while for low SNR observations, the AV-align method is lower in WER. The reason might be that under very low SNR conditions, the visual stream in the AV-cross model learns the synchronization information from the audio stream, but also learns the high intensity noise. The reduction in the  modal information gap makes the model pay more attention to the visual stream than the AV-align and the AV-concat, which amplifies the effect of the audio noise and leads to a reduction in robustness. In addition, it is worth mentioning that the WER of the AV-cross enhanced Late-fusion and Middle-fusion for -5 dB babble noise reaches up to 77.38\% and 68.35\%, respectively, while the WER of Early-fusion is only 57.94\%, i.e., as optimal as the baseline system. This confirms the benefit of using the proposed deep audio-visual encoding method.
%{\color{red}Overall, we can conclude that the AV-align based audio-visual fusion can effectively improve the performance of robust speech recognition. Specifically, the proposed AV-align enhanced Early-fusion method leads to an absolute improvement of 0.55\% and 4.51\% on average} compared to the baseline system under the clean and babble noisy conditions, respectively.
Moreover, the proposed AV-align enhanced Early fusion method can decrease the WER by 0.55\% and 4.51\% on average compared to the baseline system under the clean and babble noisy conditions, respectively. Therefore, we can conclude that the AV-align based audio-visual fusion enables a performance improvement for robust speech recognition.

In the case of the unseen buccaneer2 noise case, the obtained recognition performance is shown in Table~\ref{table:wers-buccaneer2}. This is a similar comparison to the case of the babble noise. Clearly, the AV-align enhanced Early-fusion achieves the best performance (i.e., 17.49\% WER averaged over different SNR levels) and reduces the WER by 4.61\% compared to the baseline system. More importantly, it is shown that the proposed attentive enhanced audio-visual encoding method can be generalized to the unseen noisy scenarios.

\section{Conclusions}
\label{section6}
In this work, we proposed an attentive enhanced audio-visual encoding method for robust speech recognition, where the audio-visual fusion is performed using an embedded fusion block within the encoding module. Due to the joint audio-visual encoding, the information contained in the two streams can be more completely modeled. Experimental results showed that the proposed AVSR model can sufficiently combine the two modalities and thus improve the speech recognition performance under both the clean and noisy environments. %The proposed method is helpful for the research on AVSR and the development of industrial product.
For the proposed method,  the extension of integrating audio-visual fusion into the encoding process  to other multi-modal tasks is straightforward. %We believe that it will still get a good performance since the information between different modalities can be deeply modeled.
Notably, it is interesting that the AV-cross method, which uses two-way interactions based on the MHA mechanism, achieves a lower WER for high SNRs, while its performance drops rapidly with a lower SNR. In the future, we will investigate to improve the robustness of the AV-cross method under low SNR conditions.

\end{document}